# TIME-OF-FLIGHT ENERGY MEASUREMENTS WITH BPMs *

A. Shemyakin†, J. Kuharik, R. Sharankova
Fermilab, Batavia, IL 60510, USA

*Abstract*

The energy of a bunched non-relativistic ion beam can be deduced from measuring the beam phases in neighbouring Beam Position Monitors (BPMs). This report discusses implementation of such a procedure at the PIP-II H- linac being constructed at Fermilab. The case when the flight time between BPMs is longer than the period of BPM frequency is considered in more detail. When absolute BPM phase calibration is not available, the BPM phase information can be used to track deviations of the beam energy from the desired value. Such "energy deviation" parameter is operationally implemented at the transfer line between 400 MeV Linac and the Booster, and its analog is expected to be used in the transfer line from PIP-II as well.

## INTRODUCTION

The PIP-II linac being constructed at Fermilab will accelerate H- ions up to 800 MeV [1]. At this energy, the ions are not highly relativistic, $\gamma = 1.85$, where $\gamma = \left(1 - \frac{v^2}{c^2}\right)^{-\frac{1}{2}}$, $v$ is the ion velocity, and $c$ is the speed of light. In this case, the beam energy can be calculated by measuring the flight time between BPMs in drift space. Such technique, commonly referred to as Time-of-Flight (ToF) measurement, is widely used in ion accelerators (see, for example, Ref. [2]). Typically, the BPMs report the phase offset of the bunch centroids with respect to an RF reference. For absolute energy measurements, the delays in cables and electronics need to be calibrated, as it is planned for PIP-II. If such calibration is not available but the drifts in the delays are low enough, one can use the BPM phase signals to track deviations of the beam energy from a reference. This method is implemented at Fermilab's transfer line between Linac and Booster.

## ABSOLUTE ENERGY MEASUREMENTS

The PIP-II linac design includes an instrumentation RF reference line along the linac (similar to [2], [3]) that allows absolute phase calibration of BPMs. For the ToF measurement, one or several RF cavities are turned off creating a drift space in the longitudinal plane, and BPM phase signals are measured. In most cases at PIP-II, the flight time is larger than the BPM RF period:

$$2\pi f \frac{L_1}{v} = \Delta\varphi_1 + N_1 2\pi, \quad (1)$$

where $\Delta\varphi_1$ is the fractional phase difference between the BPM phases, $L_1$ is the distance between the BPMs, $f$ is the BPM frequency (162.5 MHz for PIP-II), and $N_1$ is the integer number of periods. A procedure that includes detection of the pulse fronts to eliminate the uncertainty with $N_1$ is being discussed but has not been implemented.

The uncertainty in $N_1$ corresponds to a large uncertainty in the kinetic energy. As an example, Table 1 shows the relative energy error $\Delta E_k/E_k$ associated with assumption of $N_1$ greater by 1 than the $N_1$ actual value for the first two BPMs in the beginning of each PIP-II linac section. The uncertainty is resolved if the initial energy is known a priori with accuracy better than 10%.

Table 1: Energy Uncertainties and Energy Errors

| BPM location (section) | Initial $E_k$ (MeV) | $\Delta E_k/E_k$ (%) | $\delta E_k/E_k$ (%) |
|---|---|---|---|
| MEBT | 2.1 | 22.7 | 0.08 |
| HWR | 2.1 | 28.2 | 0.10 |
| SSR1 | 10.3 | 32.8 | 0.12 |
| SSR2 | 32 | 32.7 | 0.12 |
| LB | 177.6 | 29.0 | 0.11 |
| HB | 516 | 34.2 | 0.14 |
| Linac energy | 800 | 41.7 | 0.19 |

The uncertainty can be resolved without assumptions about the energy if three BPMs are available in the drift space, with some restrictions on the distances $L_1$ and $L_2$ between neighboring BPMs. Similar to Eq. (1),

$$2\pi f \frac{L_2}{v} = \Delta\varphi_2 + N_2 2\pi. \quad (2)$$

From Eq. (1) and Eq.(2),

$$N_2 = N_1 \frac{L_2}{L_1} + \frac{1}{2\pi}\left(\Delta\varphi_1 \frac{L_2}{L_1} - \Delta\varphi_2\right) \quad (3)$$

To determine the energy, one can use the requirement that both $N_1$ and $N_2$ are integer as follows.

1. Calculate for several integers $k$
$$M_k = k\frac{L_2}{L_1} + \frac{1}{2\pi}\left(\Delta\varphi_1 \frac{L_2}{L_1} - \Delta\varphi_2\right) \quad (4)$$
2. Calculate the fractional part of $M_k$
$$R_k = |M_k - integer[M_k]| \quad (5)$$
3. Choose $k^*$ that gives the minimum of $R_k$ since $N_2$ must be an integer. For the ideal measurement, $R_{k^*}$ is zero, and $N_1 = k^*$.
4. Calculate the velocity from Eq.(1)
$$v = \frac{2\pi f L_1}{\Delta\varphi_1 + 2\pi N_1} \quad (6)$$
and then the kinetic energy $E_k = (\gamma - 1)Mc^2$, where $M$ is the ion mass.

For demonstration purposes, the procedure was simulated at the transition between SSR1-2 and SSR2-1 cryomodules. To create a drift space, the last SSR1-2 cavity and the first two SSR2-1 cavities are turned off, and distances between consecutive BPMs are $L_1 = 1.785\ m$ and $L_2 = 2.25\ m$. The values of fractional phases differences

---

$\Delta\varphi_1$ and $\Delta\varphi_2$ were calculated using the expected value of the energy in this location. The result of calculations with Eq. (5) for different values of $k$ is shown in Fig. 1. The residual reaches zero at $k^* = 3$ as expected.

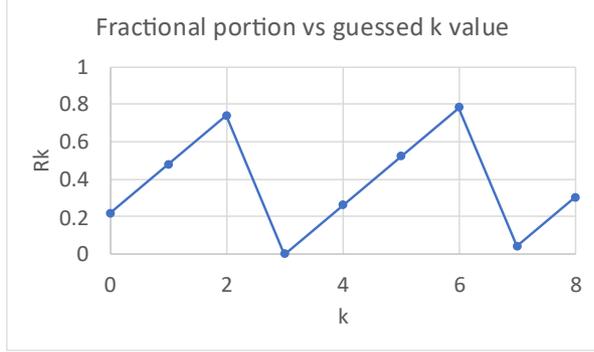

Figure 1: Values from solving Eq. (5) for $k = 0 \div 8$.

For the procedure to work, $\frac{L_2}{L_1}$ needs to be far from integer or simple rational numbers. Let's consider the case when there is a second solution of Eq. (3), a pair of integers $P_1, P_2$ in addition to $N_1, N_2$:

$$P_2 = P_1 \frac{L_2}{L_1} + \frac{1}{2\pi}\left(\Delta\varphi_1 \frac{L_2}{L_1} - \Delta\varphi_2\right) \quad (7)$$

Subtraction Eq. (3) from Eq. (7) defines the unwanted relationship between $L_2$ and $L_1$:

$$P_2 - N_2 = (P_1 - N_1)\frac{L_2}{L_1} \quad (8)$$

There are additional solutions if the ratio of distances is a rational number defined as

$$\frac{L_2}{L_1} = \frac{P_2 - N_2}{P_1 - N_1} \quad (9)$$

For example, if the ratio is a rational number with 2 in the denominator, the solution will exist for any even $k$. Fig. 1 illustrates this discussion. The ratio of distances between BPMs $\frac{L_2}{L_1} = 1.26$, which is close to the rational number 5/4. As a result, $R_k$ has a minimum every 4th integer but only one corresponding to the true value gives exact zero.

Dealing with real measurements brings a measurement error of the BPM phase measurements that can be represented by adding a component $\delta\varphi \ll 2\pi$ into the bracket of Eq. (4). The residual of the correct solution becomes $R_{k^*} = \frac{|\delta\varphi|}{2\pi}$. To be distinguishable from a different combination $(P_1, P_2)$, this residual needs to be by several times (let's say, 3) larger the residual $R_p$ at those values, so the border is $R_p = 3\frac{|\delta\varphi|}{2\pi}$. One can consider the case when the ratio $\frac{L_2}{L_1}$ is close to an integer $K$, $\frac{L_2}{L_1} = K + \Delta$, where $\Delta \ll 1$. In this case, the residual is calculated from Eq. (5) as

$$R_p = \left|(P_1 - N_1)\Delta + \frac{\delta\varphi}{2\pi}\right| \quad (10)$$

The uncertainty of the case with two BPMs can be resolved by a 3rd BPM if the difference in the length ratio from an integer is large enough to bring $R_p$ above the border value in the case of opposite signs inside the brackets of the final expression in Eq. (10) and $|P_1 - N_1| = 1$:

$$|\Delta| \geq 4\frac{|\delta\varphi|}{2\pi} \quad (12)$$

Assuming the phase measurement accuracy of 1°, the border value $|\Delta| \approx 1\%$. If the distance between two BPMs is ~2 m, adding the third BPM helps if its separation is by at least 20 mm different.

After resolving the $N_1$ value, the error $\delta E_k$ of the absolute energy measured with ToF is determined by error in knowing the distance between BPMs $\delta L_1$ and phase measurement error $\delta\varphi$:

$$\frac{\delta E_k}{E_k} = \gamma(\gamma + 1)\sqrt{\left(\frac{\delta L_1}{L_1}\right)^2 + \left(\delta\varphi \frac{\beta c}{2\pi f L_1}\right)^2}. \quad (13)$$

As an example, values estimated for $\delta L_1 = 0.1$ mm and $\delta\varphi = 1°$ (of 162.5 MHz) are shown in Table 1. Note that if all BPMs downstream of the linac that have absolute calibrations are used, the energy error drops to 0.04%.

## RELATIVE ENERGY MEASUREMENTS

If the absolute phase calibration is not available, the changes in BPM phases over time can still deliver an important piece of information about deviation of the beam energy from its reference value established by other means. It was tested in the beam line transferring 401 MeV H-beam from the Linac [4] to the Booster. This 64 m line has 20 BPMs that report beam phase counted from a reference without calibration of cable and electronics delays.

The energy of the beam injected into the Booster as well as its momentum spread are tuned by adjusting of the phase of last accelerating cavity of the Linac (Module 7) and parameters of so-called Debuncher cavity located in the last third of the transfer line. Deviations of the BPM phase readings from the reference values in the region between Module 7 and Debuncher give the beam energy deviation at the exit of the Linac, and the region between Debuncher and injection point does the same for the injection energy. Since the energy is constant over each region, the phase shift $\delta\varphi_i$ reported by $i$-th BPM changes linearly with the distance $z_i$ and energy deviation $\delta E_k$:

$$\delta\varphi_i = \frac{2\pi f}{v}z_i \frac{\delta E_k}{E_k}\frac{1}{\gamma(\gamma + 1)} \quad (14)$$

For the nominal energy and BPM reporting frequency of 402.5 MHz, Eq. (14) gives the coefficient of 0.488 deg/m/MeV. Strictly speaking, Eq. (14) is valid only for a straight line. While the 400 MeV line bends in both planes, estimations show that the corresponding corrections are below accuracy of knowledge of BPM positions.

To verify connections of BPMs, the response of BPM phases to the change of the Linac energy was recorded. The measurement was performed by changing the phase of the Module 7 RF from pulse to pulse in sinusoidal manner with 0.4 degrees (of 805 MHz, which is the cavity resonant frequency) amplitude and recording responses of BPM phases at the excitation frequency (~ 1 Hz). More details about this technique can be found in Ref. [5]. The resulting amplitudes of BPM phase oscillations vs BPM location from the end of the Linac are the shown in Fig. 2. The data in each regions fit well to linear dependence. According to Eq. (14), the energy change corresponding to this 0.4 degree phase change of Module 7 is 54 keV at the exit of the Linac and 33 keV at the injection.

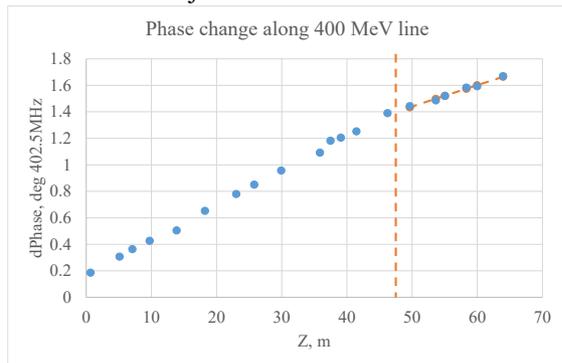

Figure 2: Shift in BPM phases along the 400 MeV line as a function of BPM distance from the Linac exit. The red line indicates the position of the Debuncher.

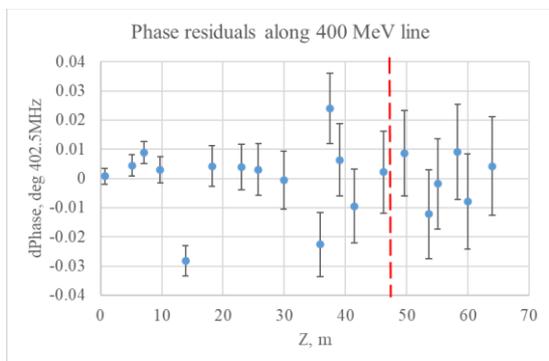

Figure 3. Deviation of measured BPM phase responses from linear fits along the line.

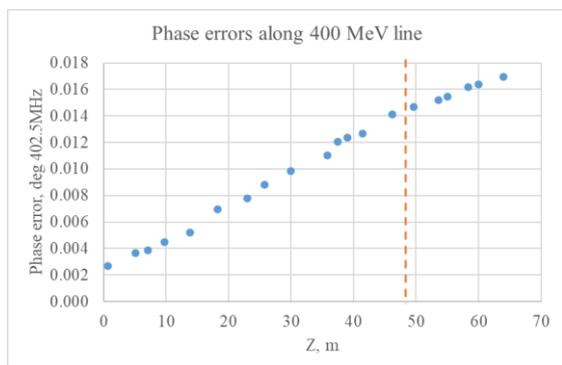

Figure 4: Statistical rms error of the phase response measurements along the line.

Deviation of the measurement points from the linear fits in each region are shown in Fig. 3. The error bars show statistical rms error $\sigma_{\varphi_i}$. Most points are within 1-2 $\sigma_{\varphi_i}$. The values of $\sigma_{\varphi_i}$ also exhibit linear dependence on the distance from the last cavity (Fig. 4). We conclude that most of the noise comes from the pulse-to-pulse Linac energy jitter of 0.5 keV rms.

With the results presented in Fig. 2-4 appearing consistent, so-called "energy deviation parameters" were established. The energy deviation is calculated by a continuously running script that updates a 10 second rolling average for each BPM with new phase readings at 1 second intervals. Reference values are then subtracted from the rolling averages. The reference values are obtained during a period when the accelerators are well tuned and stable. The difference between the measured value averages and the reference values is fit to a linear dependence on the BPM location in each region. The slope of the resulting fit is divided by the coefficient 0.488 deg/m/MeV mentioned above, and the results are assigned to two parameters in the control system. The parameters are found useful in studies and in daily tuning. Figure 5 shows another example of the energy parameter and its correlation to Booster injection losses.

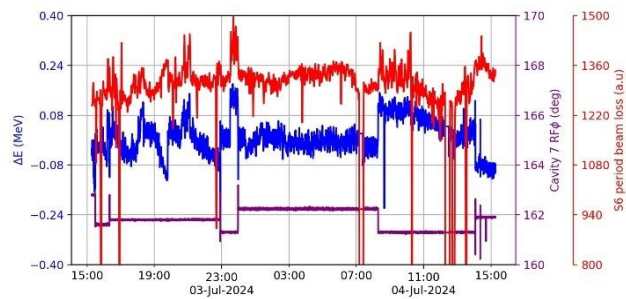

Figure 5: Booster injection loss vs. energy deviation parameter and last Linac cavity phase setting over a period of 24h.

Attempts to use the parameters in long time scales are found less successful. It is illustrated by Fig. 6, where two sets of BPM phase readings are shown with subtraction of the reference recorded 2 months before (on 3-May-24). For several BPMs, deviations from a linear fit are much larger than the statistical errors. The change in the fitted slope can be attributed to either an upstream energy change, a drift in the RF reference line due to thermal changes, or a combination of both, with the latter being the most likely as we see effects on Booster injection efficiency that roughly follow the change in the energy parameter.

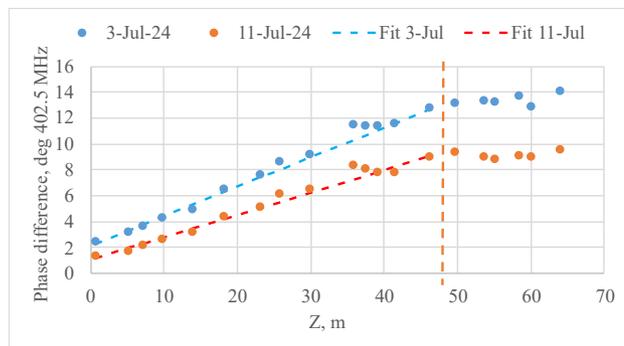

Figure 6: Deviations of BPM phase readings from the reference recorded two months prior.

## ACKNOWLEDGMENT

The authors are thankful to N. Eddy for useful discussions.